\documentstyle[12pt,aasms4]{article}

\lefthead{Hammer et al.}
\righthead{Lensing properties of the MS2137-2353 core}

\begin{document}

\title{Detailed lensing properties of the MS2137-2353 core\\ and 
reconstruction of sources\altaffilmark{1}}

\author{F. Hammer and P. Teyssandier}
\affil{DAEC, Observatoire de Paris-Meudon, 92195 Meudon, France}

\author{E. J. Shaya}
\affil{University of Maryland, Dept. of Physics, College
Park, MD 20742}

\author{I. M. Gioia\altaffilmark{2} and G. A. Luppino}
\affil{IfA, University of Hawaii, USA}

\and

\author{O. Le F\`evre}
\affil{DAEC, Observatoire de Paris-Meudon, 92195 Meudon, France}


\altaffiltext{1}{Based on observations made with the NASA/ESA Hubble
Space Telescope, obtained at the Space Telescope Science Institute,
which is operated by AURA under NASA contract NAS5-26555}

\altaffiltext{2}{Also Istituto di Radioastronomia del CNR,
Via Gobetti 101, I-40129, Bologna, ITALY}



\begin{abstract}
A deep HST image of the MS 2137-2353 core has revealed detailed
morphological structures in two arc systems, which are modelled and
well reproduced after a complete analysis of the lensing properties of
the dark matter component. Latter could have a simple elliptical mass 
distribution with ellipticity and angular orientation similar to those of the
 visible and X-ray light, which suggests that the MS 2137-2353 is a relaxed cluster at z=0.313. The predicted density profile ($\rho$ $\sim$
 $r^{-1.56\pm0.1}$ with $r_{c}$ $\le$ 22.5$h_{50}^{-1}$ kpc) within
 150$h_{50}^{-1}$ kpc implies increasing M/L ratio with the radius, and could
 be in agreement with predictions from standard CDM simulations. 

At least two faint sources (unlensed magnitude, R=23.9 and 26,
respectively) are aligned with the cluster core and are responsible of the arc
 systems. They have been reconstructed with details as small as 0".02 (or 
160$h_{50}^{-1}$ pc in the source assumed at z= 1), one could be a 
nearly edge-on barred spiral
 galaxy, and the other has a more complex morphology, which could be related to
 a close interacting pair and/or to dust. They show strong signs of star
 formation indicated by compact HII regions well off their center. The 
observation of giant luminous arcs by HST could even resolve the size of giant HII 
regions at z$\sim$ 1.    

\end{abstract}






%

\section{Introduction}

A large number of giant luminous arcs have been discovered in the cores of rich
 clusters of galaxies (e.g. Fort and Mellier, 1994). They provide the best 
estimates of the mass within few hundred of kpc of the cluster center. However, 
ground based images generally do not allow one to resolve the arc widths and 
structures, leading to a considerable uncertainty on both the shape of the 
density profile and the magnification factor (Hammer, 1991). Statistical 
approaches partially remove the uncertainty on the density profile and the 
number of arcs discovered is consistent with very small core radii ($r_{c} <$ 50 $h_{50}^{-1}$ kpc) in most of the arc clusters (Wu and Hammer, 1993). This is 
supported by the discovery of the so-called radial arcs, which require small
 core radius for the lensing cluster (Mellier et al, 1993).

There is an increasing interest to understand the dynamics of the most inner
 part of the clusters. In the framework of the 
standard CDM cosmogony, N-body simulations have predicted singular density 
profiles, approaching the halo center with $\rho$ $\sim$ $r^{-1}$ (Navarro et 
al, 1995; Tormen et al, 1996). It has been argued that the presence of radial 
arcs in cluster cores was consistent with such "universal" dark-matter halo 
profile (Baltermann, 1996) as well as with isothermal mass profile ($\rho$ 
$\sim$ $r^{-2}$) with small and definite core.

The 0".1 pixel size of the WFPC2 of the Hubble Space Telescope allows the 
resolution of most of the giant luminous arcs, which leads to considerable 
constraints on both the cluster core mass distribution and on the source 
morphology. In this letter we analyse the lensing properties of the MS 2137-2353
 core in which two multiple image systems have been discovered so far (Fort 
et al, 1992), and which have been observed by the HST (Gioia et al, 1996a).  
 Throughout the paper, values of $H_{0}$=50 and $q_{0}$=0.5 are assumed.            

\section{Image analysis}

\subsection{Arc templates}

Before modelling, arc images should be cleaned 
from the contamination due to the cluster galaxy light, especially the brightest cluster galaxy G1 and the galaxy G7 which is superimposed to the edge of the 
giant arc. Arc template images used to feed the lensing model have been also 
limited above the threshold of 1$\sigma$ above the average background noise. The standard package STSDAS (task ellipse) in IRAF has been used to model the 
contaminating galaxies G1, G7 and the four small galaxies lying within the 
envelope of G1. Figure 1 shows the cluster core after removing the background 
and the contaminating galaxies G1 and G7. Within the envelope of G1, the 
residual noise is 1.5 to 3 times higher than the background sky noise.  We
 find no evidence for a fifth image which could be associated to 
the arc system (A0, A2 and A4). The radial arc (AR) extends from 3".4 to 
7".0 from the mass center and is surrounded by a rather complex structure.     

\subsection{Optical properties of the cluster}

The brightest cluster galaxy G1 is  well fitted by an ellipse with ellipticity varying from 0.12$\pm$0.02 in the very center (r=2") to 0.16$\pm$0.02 beyond
 r=6" (or 34$h_{50}^{-1}$ kpc) and P.A.= 47.5$\pm$5 degrees (the header keyword 
ORIENTAT of the HST image is 142.5). Both ellipticity and P.A. of G1 are in 
rather good agreement with values derived from the X-ray gas 
($\epsilon$=0.13$\pm$0.02 and P.A.=65$\pm$15 degrees, Gioia et al, 1996a). Its 
surface brightness profile does not follow a $r^{1/4}$ law, while it shows a 
rather steep profile beyond r=3" ($\beta$=1.6 for $\sigma$$\sim$ $( 1 + 
(r/r_{c})^{2})^{0.5-\beta}$). The latter number is derived from the output of 
the task ellipse (package STSDAS). The HST image quality reveals that the galaxy G7 is a nearly edge-on spiral galaxy (axis ratio=0.54) with a nicely defined 
bulge (it is probably an Sa).

\section{Lensing model and methodology}

Gravitational bending angles can be computed for various sets of mass density 
profiles in the code AIGLE (Astronomical Instrument for Gravitational Lensing 
Experiments). It is an up-dated version of the code written by Hammer and Rigaut (1989) and includes inversion of the lensing equation through the Schramm and 
Kayser (1987) method. Elliptical lenses are assumed to follow the homoeidal mass distribution (see Schramm, 1990) and the projected isodensities are concentric 
ellipses, with various density profiles. Mass distribution are used as input of 
the code rather than potential. Several 
techniques have been implemented to optimise the modelling, including $\chi^{2}$ tests for the comparison between the arc fit and the arc template, and self 
consistency and coherence tests for the source reconstruction in the case of 
multiple image configuration (see e.g Wallington et al, 1995). For a system 
with four images of the same source, the 
basic methodology to model the arcs consists of the following three steps : (i)  to invert the lensing equation for a given sub-structure (generally the 
brightest knot) of the source, to compare the fit to the observations and to 
optimise it, and hence, to investigate the degeneracy of the parameters; (ii) to reconstruct the source for each the 4 individual images and test the consistency between the 4 reconstructed source images (RSIs); (iii) to combine the RSIs and 
to reconstruct the source of the system, and by using the most self consistent 
reconstructions, to invert again the lensing equation and to test the result by 
comparison with the arc templates. The latter test securely ensures that the 
reconstructed source is not providing extra-images which are not observed.





\section{Model of the arcs}

We have assumed a single $\beta$ profile model ($\rho$=$\rho_{0}$ $(1 + 
(r/r_{c})^{2})^{-\beta}$) for the cluster core, and have investigated the space 
of the mass parameters (P.A., $\epsilon$, $r_{c}$, $\beta$ and $\sigma^{2}$) to 
reproduce the arc location. Source reconstruction is then derived and self 
consistency tests are applied. We finally consider perturbations from the 
secondary deflectors (such as G1 and G7) and discuss them.
Examination of the arc templates confirms the identification of one multiple
image system which is composed of 4 images (A01, A02, A2 and A4) of a single
 source (S1). No evidence is found for a fifth image near the center of G1, 
which suggests that it is actually overfocused (low magnification factor).
The other system is made of 3 images (A6 and AR) of a second
 source (S2), the radial arc being a complex mix of two images. 

\subsection{Images A01, A02, A2 and A4 of the S1 source}

For a large variety of sets of mass parameters, we have been able to reproduce 
the location of the brightest knot in each of the four images, within 1 pixel 
accuracy. The P.A. of the lens major axis is well constrained within 41-46 
degrees, in rather good agreement with both optical and X-ray light. There is an obvious degeneracy between the other parameters, which is illustrated in Figure 
2. Assuming a value for $r_{c}$ and $\epsilon$ ($>$0.06) generally provide a 
solution which set $\beta$ and the mass (proportional to the square of the line 
of sight velocity dispersion, $\sigma_{los}$). Values for the latter do not 
differ from the ones found by Mellier et al (1993), for $\beta$=1 and $r_{c}$=7" (for which we find $\sigma_{los}$=1216$ km s^{-1}$ if the source is assumed at 
z=1), and differences between mass estimations from lensing and from X-ray 
measurements are discussed by Gioia et al (1996a). We have mapped the ($r_{c}$, 
$\epsilon$) plane assuming several values for $\beta$. For each set of 
parameters ($r_{c}$, $\epsilon$ and $\beta$), we have calculated the 4 
reconstructed source images (RSIs) corresponding to the 4 arcs (A01, A02, A2 
and A4), assuming a pixel size in the source plane of 0".02 (Figure 3). It
 results that the arc A0 is a double image of only a fraction of the source,
 because the source is superimposed on to the diamond shape caustic line. This
 is a good illustration of the predictive model for an elliptical lens made 
by Bourassa and Kantowski (1975). To test the consistency (or similarity) 
between the 4 RSIs, we have calculated the angular distances from the brightest 
knot to 2 other knots found in the RSI of A01 and A02, and to 4 other knots 
found in the RSI of A2 and A4 (see Figure 3). The self similarity parameter is 
defined as:

SSP = $\Sigma_{k}$ $\Sigma_{i>j} | d(A_{i},k) / d(A_{j},k) -1 | / 6$, 

where the summations have been done on the knots (k), the RSIs (i), and the 
angular distance between the brightest knot and the knot (k) in the RSI (i) is 
$d(A_{i},k)$.
Values of SSP are indicated in Figure 2, and typical error bars are $\pm$ 0.08. 
It takes minimal values ($\sim$ 0.3) for $\beta$ ranging between 0.6 to 0.85, 
leading to ellipticities ranging from 0.06 to 0.18 and core radius lower than 
8".  $\beta$=1 profiles imply too large sizes for the RSIs of A4, A01 and A02 compared to the RSI of A2. Steeper mass profiles ($\beta$=1.5) 
provide even more implausible solutions. 
Figure 3 shows one of the best reconstruction which corresponds to one of the 
minimum value for the self similarity parameter (SSP =0.35 with $\beta$=0.8, 
$r_{c}$=2 and $\epsilon$=0.151). The reconstructed source images (RSIs) show 
very nice consistency, especially when one accounts for the fact that the RSIs 
of A01 and A02 are only sampling a small (top-left) part of the source. Both the noise and artifacts are damped when the RSIs are combined, providing a detailed 
image of the actual source (S1) of this arc system (Figure 3, middle panel). 
Other 
best reconstructions of the source present very similar morphology and only the 
source size is affected, which corresponds to variation in the magnification 
factor. The simulated arcs can be rebuilt by re-imaging the reconstructed source by the model, and can be compared to the observed one (Figure 1). Again, the 
similarity between the simulation and the actual arc images is very good, even 
in the small details. 

There are however some discrepancies between the 4 reconstructed source images (RSIs). Also our investigation of a large fraction of the parameter space  has not provided a solution with a self similarity parameter (SSP) very close to 0. We believe that several reasons can explain these residual discrepancies: (i) the mass profile is too naive and $\beta$ is varying with the radius; (ii) there are perturbations by secondary lenses (cluster galaxies); (iii) there might be an additional elliptical component with a different orientation than the main one; (iv) the compactness of some knots (such as the brightest one), provides artificial elliptical shapes in the RSIs; (v) we suspect that one faint knot in the arc A2 is a contaminating object. We have tested the influence of G7 as a secondary lens. Acceptable mass values for G7 (which corresponds to $\sigma$=120$km s^{-1}$ for an isothermal profile) are equal or below 1\% of the cluster core mass, and slightly affect the predicted parameters for the cluster mass distribution (for example it decreases the ellipticity by $\sim$ 10\%). The problem associated with larger values for the G7 mass, is that they imply too large sizes for the reconstructed source image (RSI) of A02 (compared to the RSI of A2), and degrade significantly the source reconstruction ($\sigma >$ 150 $km s^{-1}$ for G7 is rejected at a 3$\sigma$ level). 

\subsection{Images A6 and AR (radial arc) of the source S2}

To model this arc system, we have tested each of the parameter sets which 
reproduce the arc location in the formerly discussed arc system (A0, A2 and A4, 
see Figure 2). The extension of the radial arc very close to the mass center 
(3".4) implies very small values for the core radius $r_{c}$ ($<$ 3".5 for 
$\beta$=0.8). Figure 2 shows the available range of parameters which can 
reproduce the radial arc location (left side of the dotted line). For this 
range of parameters, 
the source (of A6 and AR) should have a redshift very similar to that of the source (S1) of A0, A2 and A4 (assuming z=1 for the latter implies 1 $\le$ z 
$\le$  1.03 for the former). The source (S2) has a morphology which can be
 suggestive of a nearly edge-on barred spiral (Figure 3) with rather 
asymetric arms. The radial
 arc is most likely a blended image of the bar, almost aligned towards the mass 
center (as the radial arc). A re-imaging of the source S2 is 
presented in Figure 1, and well reproduces the length and the location of 
the radial arc. However it cannot reproduce the bright knot 1".4 above AR, 
which should be related to another system. We found that this knot could be an additional image of the A5 source (S3) assumed at redshift slightly below that
 of S2.

\section{Discussion}

\subsection{Mass distribution in the MS 2137-2353 core}

Figure 2 summarizes the mass parameter predictions provided by the fit of the 
two arcs systems in the HST image. To reproduce the arc system (A0, A2 and A4) a flat density profile is required ($\beta<$ 0.85), while the presence of the 
radial arc very near to the mass center implies a very small core radius ($r_{c} \le$3.5 for $\beta$=0.8). The only possibility to fit the radial arc with a 
$\beta$=1 profile implies large ellipticities ($\epsilon$ $>$ 0.22) and gives 
highly inconsistent source reconstruction from the 4 image (A01, A02, A2 and A4) system (SSP$\ge$0.81). This mass profile for the MS 2137-2353 core can be 
rejected at a 8$\sigma$ level.

This analysis provides a considerable reduction of the available volume of the  
space of parameters which can fit the arc observations, especially when compared to results based on  ground-based observations (see Figure 2). The solution 
found by Mellier et al (1993) ($\beta$=1 and $r_{c}$=7") is inconsistent with 
both arc systems. This is simply related to the fact that, from the ground, the 
radial arc was found far less extended towards the mass center than in the HST 
image, and that no morphological information can be obtained from unresolved 
arcs. The range of parameters which provide reasonable fit of the two arc 
systems are 0.6 $<$ $\beta$ $<$0.85, 0.06$<$ $\epsilon$ $<$ 0.18 and $r_{c}$ $<$ 4". It is likely that the available range for the parameter is even smaller, if 
one assumes that the ellipticity of the dark matter matches well the ellipticity of both the visible and X-ray matter (0.12 $<$ $\epsilon$ $<$ 0.17). If true, 
the range of mass parameters (described in Figure 2 by a shaded area) is:
$\beta$ =0.78 $\pm$0.05, $\epsilon$ = 0.145 $\pm$0.025 and $r_{c} <$ 4" (or 
22.5$h_{50}^{-1}$ kpc).

For this range of parameters the expected fifth image of the (A01, A02, 
A2 and A4) system is predicted overfocused (amplification factor $<$ 1), 
in agreement with the observations (Figure 1). This analysis does not depend 
on to the 
(unknown) redshift of the source, since we have treated the mass ($\sigma^{2}$) 
in the lensing equation as proportional to $D_{s}/D_{ds}$, where $D_{s}$ and 
$D_{ds}$ are the angular diameter distance of the source, and from the lens to 
the source, respectively.

\subsection{Properties of the sources}

Since the available range of mass parameters is well constrained, one can derive the properties of the source (e.g. luminosity and size) with an unprecedented 
accuracy. The magnification factor is known with an accuracy of better than 15\% for both arc systems.

The arcs A2 and A4 are images of the same source (S1) which is magnified by 
13$\pm$1.5 and by 9.2$\pm$1.0, respectively. This implies that the unlensed 
source (S1) of the (A0, A2 and A4) system would appear as a R=26.0$\pm$0.25
 galaxy, a substantial fraction of the uncertainty being related to photometric
 measurement errors. The corresponding magnification factors for the arcs A01
 and A02 reach 
very high values (46$\pm$8). We believe rather unsecure the derivation of 
redshift information for an object with such a flat spectral energy distribution
 ($B-R$= 0.4 from Fort et al, 1992). Such colors, as well as the absence of 
emission line in the optical spectrum could be associated with any intrinsically
 blue galaxy with 0.8 $<$ z $<$ 2.5, a rather usual range for R=26 galaxies. As 
a whole, the source has an axis ratio of 1.7 and is rather compact, with a major axis of 0".90$\pm$0.05, which corresponds to 7.1$\pm$0.2$h_{50}^{-1}$ kpc after 
accounting for uncertainties related to the magnification factor and to the 
redshift undetermination. Its morphology shows some resemblance with the 
reconstructed source of Cl 0024+1654 arc system (Colley et al, 1996), although 
it is far more complex. Among the 600 local galaxies taken from the compilation of the Color Atlas of Galaxies  (Wray, 1988), eight galaxies show some
 morphological similarities with the 
source of the (A0, A2 and A4) system, and are by decreasing order of 
resemblance, NGC 4038, 4618, 0337, 1385, 3256, 6221, 1559 and 3162. Latter have type ranging from T=6 to T=9, and U-V color from 0.25 to 0.6. 3 of them are young irregular systems, 2 shows irregularities 
because of dust, 2  (including NGC 4038) are the result of galaxy interactions, 
and the remaining one has a barred spiral morphology. There is little doubt 
that S1 
is a rather young and irregular system at high redshift, while there are some 
possibilities that either dust or close interaction between two galaxies are 
contributing to its appearence. The source brightest knot (on top right of the 
reconstructed source, see Figure 3) is probably an HII region, since the F702W 
filter corresponds to ultraviolet light in the rest-frame source. It is 
apparently unresolved in arc A2 and A4, while the large magnification factors in arcs A01 and A02 resolve it into two small elements separated by 0".26 (Figure 1), which correspond to 
0".015, or 120$h_{50}^{-1}$ pc in the actual source. The gravitational 
microscope can also provide us detailed informations on the size of the bright 
and giant HII regions in distant star forming galaxies.

The unlensed source (S2) of the (A5 and AR) system would appear as a 
R=23.9$\pm$0.2 galaxy which resembles to a nearly edge-on barred spiral 
(Figure 3). The lensing model predicts a very similar redshift for the sources (S1 and S2) of 
the two arcs systems ($\Delta$z$\le$0.03 at z=1). The (unlensed) projected distance between the two sources is 3".2$\pm$0".3, which corresponds to 26$\pm$3$h_{50}^{-1}$ kpc, the 
error bar accounting for the uncertainties related to the modelling and to the 
undetermination of the source redshifts. Assuming that A5 is actually related
to the source (S3) of the bright knot above AR would imply that 3 sources with 
similar redshift are within the large magnification area which has a 
radius smaller than 2" (Figure 3). They might be either interacting galaxies
 or the result of projection of galaxies lying in a large structure of
 galaxies.

\section{Conclusion}

A very simple mass distribution for the dark matter accounts for most of the 
detailed lensing properties of the two arc systems found in the core of MS 
2137-2353, after careful analysis of a deep HST image of the cluster. Within 33 
to 150$h_{50}^{-1}$ kpc from the mass center, the major axis and ellipticity of 
the dark matter component are in a good agreement with those derived from X-ray 
and visible light, while the dark matter density profile has a slope  
($\beta$=0.78$\pm$0.05) much flatter than the visible light ($\beta$=1.6). MS 
2137-2353 is probably a good example of an essentially relaxed cluster at 
z=0.313, with an increasing mass to light ratio from the very center to r$\sim$ 
150$h_{50}^{-1}$ kpc. The mass distribution profile should be associated with a 
very small or a null core radius ($r_{c} \le$ 22.5$h_{50}^{-1}$ kpc), and could 
be associated with a single power law, $\rho$ $\sim$ $r^{-1.56\pm0.1}$ within a 
150$h_{50}^{-1}$ kpc radius from the cluster center. This could be in agreement 
with the universal profile predicted by standard CDM simulations (Navarro et al, 1995; Tormen et al, 1996). This analysis also brings some support to the very 
simple analysis of a large number of arc systems by Wu and Fang (1996) who found a similar value for the slope on average, although they have neglected the 
effect of subclustering.

The relatively low value for the slope of the cluster mass density profile 
implies large magnification factors, and provide us with the opportunity to look at details as small as 0".02 in the reconstructed sources. The two sources 
associated with the two arc systems in the MS 2137-2353 core are very close in 
redshift space  and might be interacting objects. While they are high redshift 
galaxies (0.8 $<$ z $<$ 2.5), they both have morphologies which are not so 
unusual compared to that of present-day galaxies. One could be a nearly edge-on 
barred spiral, the other has a more irregular morphology, which combined 
with its blue color, 
suggests a young star-forming system in a closely interacting pair possibly 
affected by dust. Both sources also show strong peaks of emission, well off 
their centers, which are likely associated with HII regions, and hence indicate 
star formation. Magnification factors in giant arcs can reach so large values 
that HII regions can be spatially resolved by this technique.      

\acknowledgments

We thank M. C. Willaime for her help in writing the code AIGLE, X.P. Wu for 
comments on the manuscript and B. Monart for her help in comparing the arc 
source morphology to those of local ones. This work has received partial 
financial support from NASA-STScI grant GO-5402.01-93A, NASA grants NAG5-2594 
and NAG5-2914.

\clearpage

\figcaption[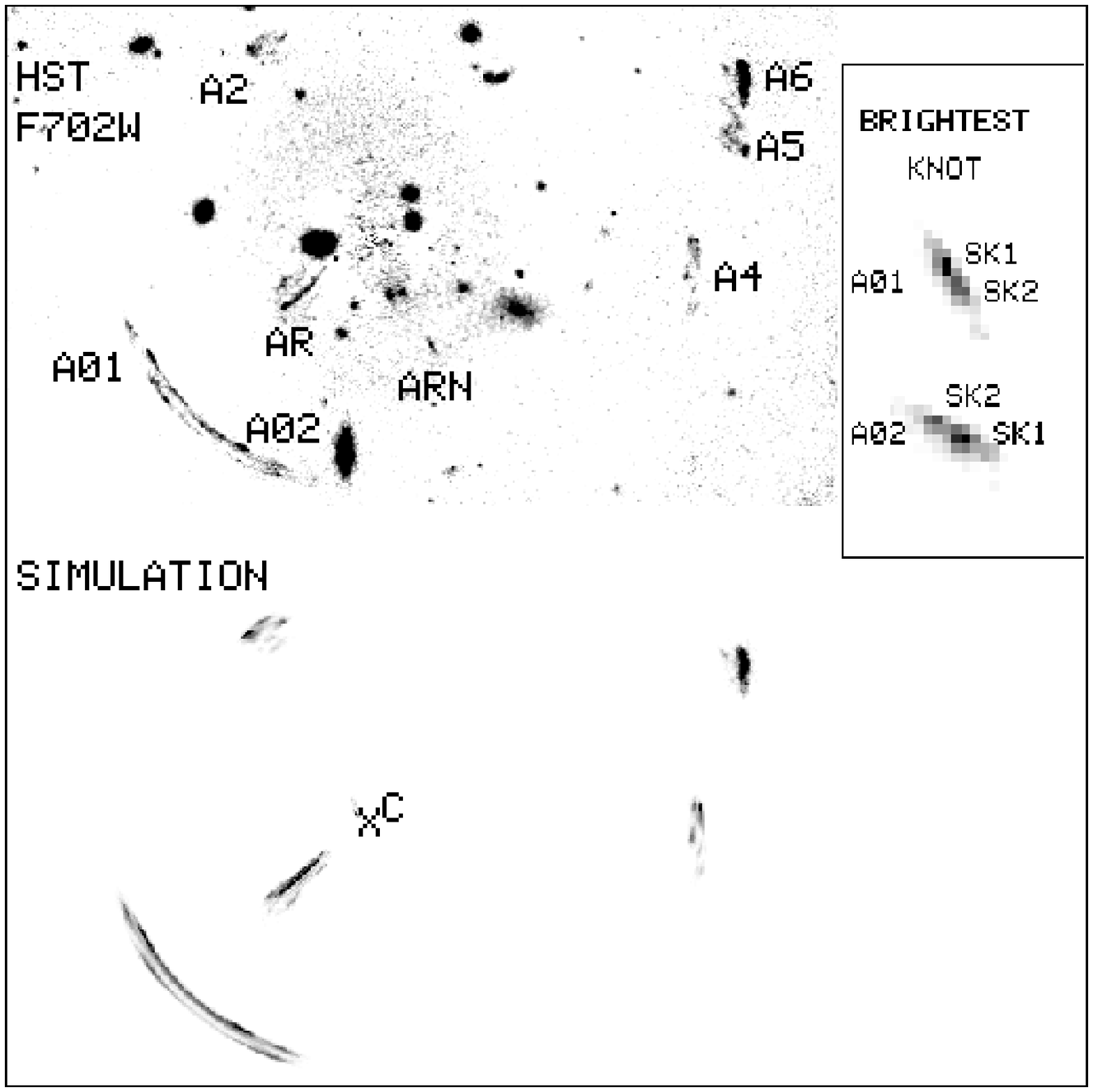]{(top): MS 2137-2353 core after subtraction of G1 and G7 
galaxies. The subtraction of G7 
reveals the complex structure of the bottom-right part of the giant arc A0. 
Arc images are labelled as in Fort et al (1992). The giant arc is made by the
 blend of two images (A01 and A02) of the source (S1), also associated with 
arcs A2
 and A4. No extra image is seen near the cluster center. The other arc system
 is made by the radial arc (AR) and A6 (source S2). The radial arc extends 
very close to the cluster center (3".4). There is also another possible radial 
arc indicated as ARN. 
(bottom): reconstruction of the arcs assuming two sources (S1 and S2) at 
nearly the same 
redshift (see text), for a model with $\beta$=0.8, $r_{c}$=2" and 
$\epsilon$=0.151. Most of the morphological details are well reproduced, with 
the noticeable exceptions of the left end of the arc A01 and of the bright knot 
above the radial arc. Latter is probably related to A5 (source S3). The fifth
 image of 
the arc system (A0, A2 and A4) is predicted too faint (near the cross
 which indicates the mass center), to be detected in the HST image.
(box in the top right): zoom of the brigthness knot in A01 and A02, 
respectively. While this knot is unresolved in A2 and in A4, the extremely
 large magnification factor in the giant arc allows to resolve it in two 
sub-knots (SK1 and SK2) in the two images. Note that it is consistent with the 
fact that A01 and A02 are reversed images (the brightest pixel is found in SK1
 in both images). The separation between the two sub-knots is 0".26 which correspond to 
0".015 in the unlensed source.\label{fig1}}

\figcaption[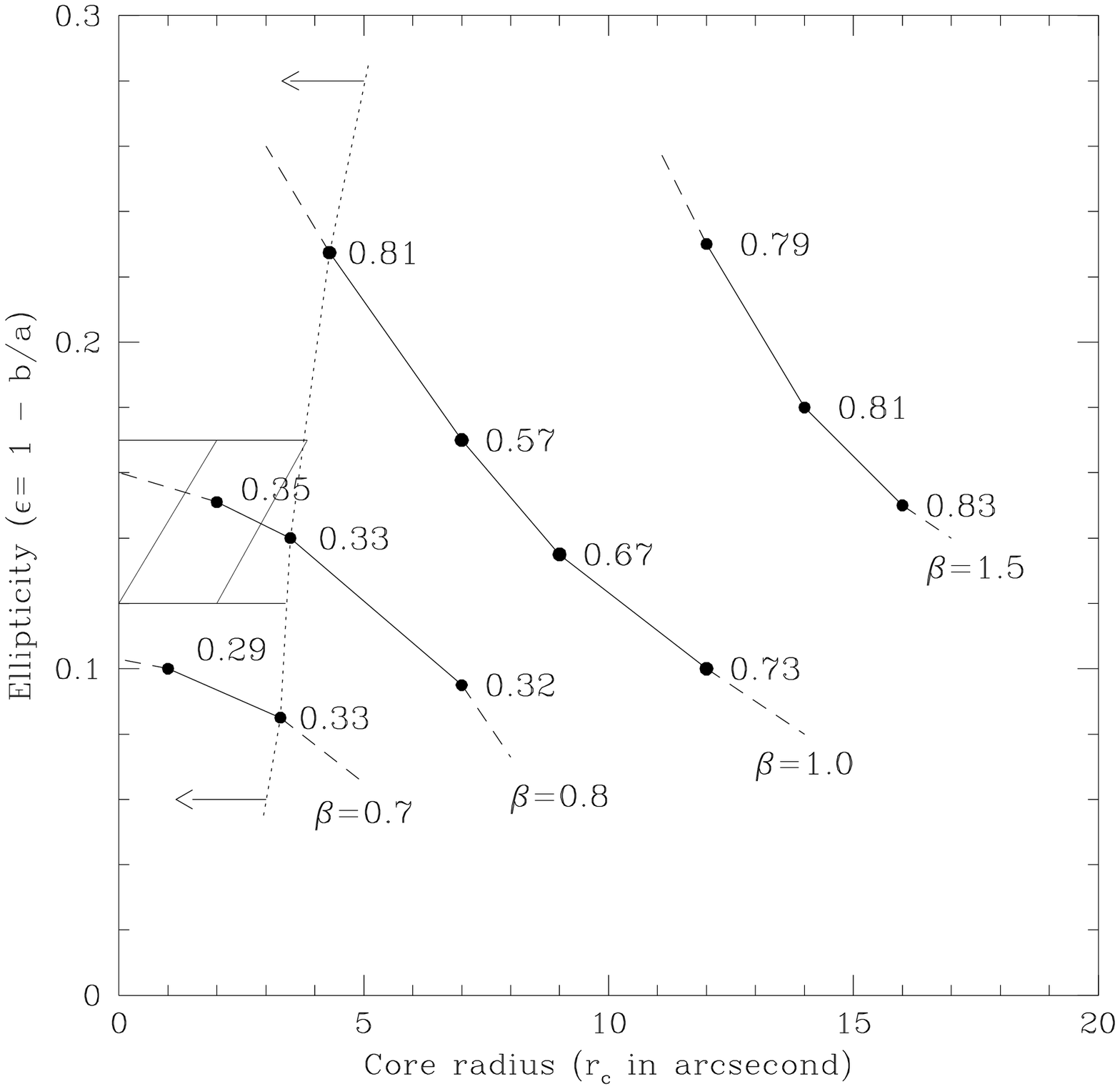]{The ($r_{c}$, $\epsilon$) fundamental plane of mass 
parameters. Each point corresponds to a fit within 1 pixel accuracy of the 
brightest knot in arcs A01, A02, A2 and A4. The values of the self consistency 
parameter (SSP) are indicated near each points. Solid lines connect the points 
for constant $\beta$. Dotted line delimits the area (indicated by arrows) of 
parameters which fit the location of the radial arc. It has been empirically 
determined from systematic tests for each set of parameters. Shaded area is the 
most likely area for the mass parameters, assuming that the ellipticity of the 
dark matter component matches those of the X-ray and visible matter. \label{fig2}}

\figcaption[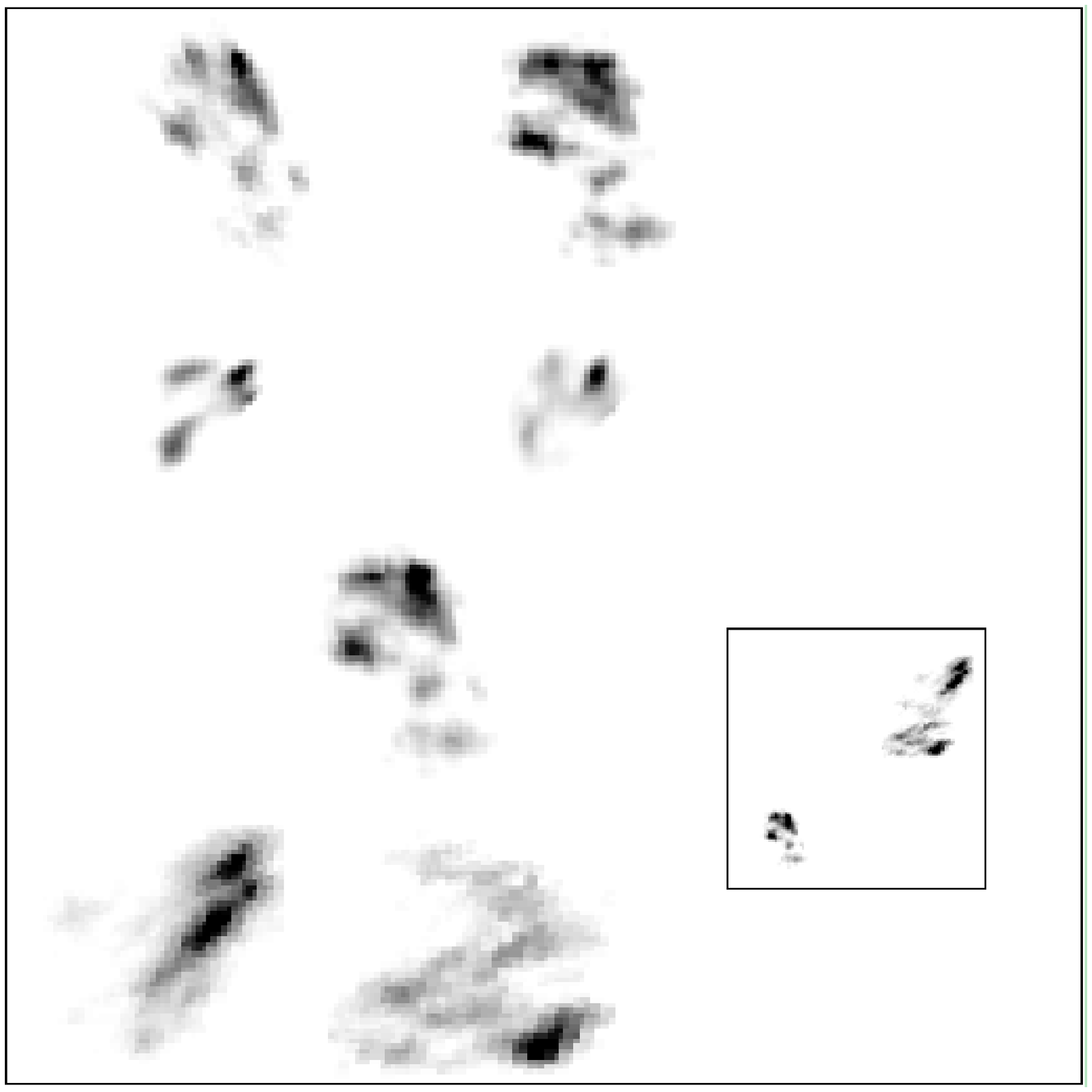]{The four top panels show the reconstructed source images 
(RSIs) corresponding to arcs A2, A4, A01 and A02, respectively (from the left to the right and from top to bottom). The corresponding reconstructed source (S1) 
is the combination  of these four RSIs, after re-scaling the intensities, and 
is shown just below 
the four top panels. The lowest panel shows the reconstructed sources (S2 and
 S3, from left to right) of arcs A6 and AR, and of A5 and the bright knot 
above the radial arc (AR). In the box at the bottom right are displayed the 
model predicted locations of the sources (S1, S2 and S3) if there were no
foreground lensing cluster on their line of sight. Size of the box is
4".\label{fig3}}

\end{document}